\documentclass[letterpaper,twocolumn,10pt]{article}
\usepackage{usenix,epsfig}

\newfont{\mycrnotice}{ptmr8t at 7pt}
\newfont{\myconfname}{ptmri8t at 7pt}

\usepackage{lastpage}
\usepackage{xspace,subfigure,multirow,upgreek}
\usepackage{color,cite,lastpage}
\usepackage[table,xcdraw]{xcolor}
\usepackage[font={small},bf]{caption}
\usepackage[hyphens]{url}
\usepackage{hyperref}

\def\testname{0}	%set 1 for weighted KS-test
\def\shownotes{1}   	% set 1 for version with author notes
\def\showedits{0} 	%set for 1 to turn edits red, 0 for them to become black.

\def\showimcedits{0} 	%set for 1 to turn edits red, 0 for them to become black.

\ifnum\testname=1

\else

\fi

\newcommand{\xref}[1]{\S\ref{#1}}

\ifnum\shownotes=1
\newcommand{\authnote}[2]{{ $\ll$\textsf{\footnotesize #1 notes: #2}$\gg$}}
\else
\newcommand{\authnote}[2]{}
\fi
\newcommand{\Pnote}[1]{{\color{purple}{\bf{\authnote{Phillipa}{#1}}}}}

\ifnum\showedits=1

\else

\fi

\ifnum\showimcedits=1
	
\else
	
\fi

\providecommand{\vs}{vs. }
\providecommand{\ie}{\emph{i.e.,} }
\providecommand{\eg}{\emph{e.g.,} }

\providecommand{\myparab}[1]{\smallskip\noindent\textbf{#1} }

\title{The Politics of Routing: Investigating the Relationship Between AS Connectivity and Internet Freedom}

\author{[Paper \#16-- \pageref{LastPage} pages]}
\author{
{\rm Rachee Singh}\\
Stony Brook University
\and
{\rm Hyungjoon Koo}\\
Stony Brook University
 \and
 {\rm Najmehalsadat Miramirkhani}\\
Stony Brook University
 \and
 {\rm Fahimeh Mirhaj}\\
Stony Brook University
 \and
 {\rm Phillipa Gill}\\
Stony Brook University
 \and
 {\rm Leman Akoglu}\\
Stony Brook University
} 

\newcounter{GraphCounter}
\newenvironment{packeditemize}{\begin{list}{$\bullet$}{\setlength{\itemsep}{0.2pt}\addtolength{\labelwidth}{-4pt}\setlength{\leftmargin}{\labelwidth}\setlength{\listparindent}{\parindent}\setlength{\parsep}{1pt}\setlength{\topsep}{0pt}}}{\end{list}}

\newcommand{\squishenum}{
   \begin{enumerate}{}
    { \setlength{\itemsep}{0pt}      \setlength{\parsep}{0pt}
      \setlength{\topsep}{3pt}       \setlength{\partopsep}{0pt}
      \setlength{\leftmargin}{1.5em} \setlength{\labelwidth}{1em}
      \setlength{\labelsep}{0.5em} } }

\newcommand{\squishlist}{
   \begin{list}{$\bullet$}
    { \setlength{\itemsep}{0pt}      \setlength{\parsep}{3pt}
      \setlength{\topsep}{3pt}       \setlength{\partopsep}{0pt}
      \setlength{\leftmargin}{1.5em} \setlength{\labelwidth}{1em}
      \setlength{\labelsep}{0.5em} } }

\newcommand{\squishlisttwo}{
   \begin{list}{$\bullet$}
    { \setlength{\itemsep}{0pt}    \setlength{\parsep}{0pt}
      \setlength{\topsep}{0pt}     \setlength{\partopsep}{0pt}
      \setlength{\leftmargin}{2em} \setlength{\labelwidth}{1.5em}
      \setlength{\labelsep}{0.5em} } }

\newcommand{\squishend}{
    \end{list}  }

\newcommand{\squishenumend}{
	\end{enumerate}	}

\begin{document}

%% COPYRIGHT STUFF
\iffalse
\conferenceinfo{SIGCOMM'11,} {August 15-19, 2011, Toronto, Ontario, Canada.}
\CopyrightYear{2011}
\crdata{978-1-4503-0797-0/11/08}
\fi

\clubpenalty=10000
\widowpenalty = 10000

\maketitle

\section*{Abstract}

The Internet's importance for promoting free and open communication has 
led to widespread crackdowns on its use in countries around the world. 
In this study, we investigate the relationship between national policies 
around freedom of speech and Internet topology in countries around the world. 
We combine techniques from network measurement and machine learning to identify 
features of Internet structure at the national level that are the best indicators 
of a country's level of freedom. We find that IP density and path lengths to other countries 
are the best indicators of a country's freedom. We also find that our methods predict freedom categories
for countries with 91\% accuracy.

%\category{C.2.2}{Computer-Communication Networks}{Network Protocols}
%%\category{C.4}{Performance of Systems}{Measurement techniques}
%
%\keywords{FILL IN}

\section{Introduction}

The Internet's role as a communication tool for activists and dissidents has led to increasing efforts on the part of nation states to restrict access and control information accessed online~\cite{arabspring}. These efforts to clamp down on Internet freedom can lead to national policies that influence the interdomain topologies of given countries (\eg Iran's policy that all networks must connect via the national telecom AS 12880~\cite{oniiran}) and the topologies in turn can make certain forms of information control easier (\eg country-wide Internet shut downs~\cite{dainottiIMC2011}). 

In this study, we consider the relationship between interdomain topology--\ie routing between autonomous systems (ASes)--and Internet freedom in countries around the globe\footnote{For simplicity, we consider ASes registered within a given country as comprising the country's AS-level graph.}. We use a combination of standard graph theoretic metrics and domain-specific features to understand how these features relate to Internet freedom, which we quantify using the Freedom House Freedom of the Press index~\cite{fpi}.\footnote{We use this index 
over the Freedom on the Net score~\cite{fni} because it has been calculated for 199 countries, whereas Freedom on the Net only covers 66 countries.} Our goal is to understand the way that online information controls can impact the network topology of different countries as well as which topologies are more likely to facilitate restrictions on Internet access. Understanding this relationship can help fill in gaps in existing data sets about Internet freedom, many of which require manual effort and measurement points within the region to compute. By looking at the interdomain topologies we can understand which countries are similar in terms of network structure and identify regions that warrant more investigation. Further, by understanding network properties that correlate with information controls we can potentially identify countries that are in a good position to perform filtering or Internet shutdowns and see the legacy effects after filtering has been repealed.
While it may seem simple, studying the interdomain topology around
the globe requires care to avoid known blind spots in existing data. We perform traceroutes using RIPE Atlas~\cite{ripeatlas} to expose additional edges in and around different countries (\xref{sec:improv_cov}). 
%We also leverage an optimized	methodology for running active measurements from each country to expose as many new edges as is possible.
We use a state-of-the-art BGP path simulator ~\cite{quicksand} which allows us to consider 
graph theoretic metrics with paths that incorporate routing policy \vs simple shortest path.
We combine our empirically derived data with existing AS-level topologies~\cite{CAIDAtopo} to characterize the interdomain topologies of countries around the globe. We consider a variety of 
structural features and domain specific features and characterize their relationship to 
free communication and find that the number of IP addresses per individual (IP density) is 
the most important predictor of Internet freedom(\xref{sec:feature_influence}).  
We leverage these features and machine learning techniques to group countries based on their interdomain topologies (\xref{sec:ml}).

\section{Internet Topology Data Sets}
\label{sec:trcrt}

In this section, we provide background on the Internet topology and 
describe our methodology for improving the fidelity of existing data sets.

\subsection{Initial Topology}

\begin{figure}[t]
\centering

\includegraphics[width=0.30\textwidth, trim={0cm 0cm 0cm 0cm},clip]{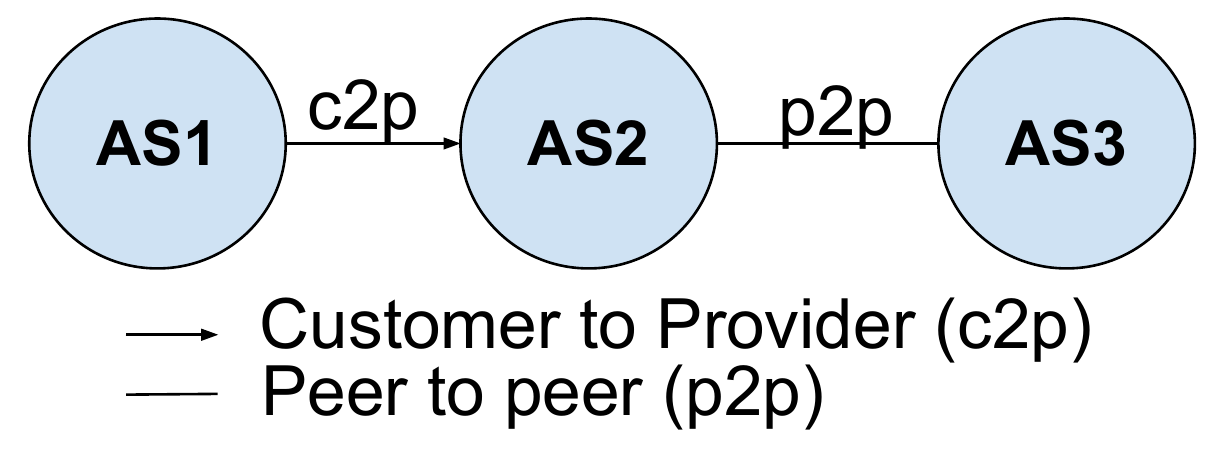}
\caption{AS relationships on the Internet}
\label{fig:topo}
\end{figure}

Central to our approach are graphs of the Internet topology of each country. These 
graphs are empirically derived and are comprised of autonomous systems (ASes) as nodes 
and connections between them as edges. An AS generally represents a network 
under the control of a single entity (\eg an ISP, university or business). Each edge is annotated with the inferred business 
relationship between the two ASes it connects (\ie who pays who). Figure~\ref{fig:topo} 
illustrates an example topology where AS 1 is a customer of AS 2 and pays it for transit 
and AS 2 and AS 3 are settlement-free peers exchanging traffic at no cost.
We start from the 
AS graph published by CAIDA~\cite{CAIDAtopo} which includes all observed ASes and is labeled 
with inferred business relationships~\cite{caidaIMC2014}. 

Since ASes may span multiple countries and even continents, we make a simplifying 
assumption and consider an AS to belong to the country it is registered in. 
We determine where each AS in the topology is registered using datasets from the 
regional Internet registries (RIRs).

\iffalse
A key challenge we faced in this work, was how to deal with known incompleteness 

The ASes registered in a country forms the set of nodes in the country's graph. We then find the connectivity between these nodes using CAIDA's AS relationship dataset. The AS relationship dataset provides pairs of ASes that are connected to each other and the business relationship between them. While the AS relationship dataset provides us a way to bootstrap the building of per country AS graphs, it is known to be incomplete. We tackle this issue by augmenting the AS graphs with AS edges inferred from traceroute measurements.

\subsection{Initial AS-level Topology}

CAIDA's AS relationship dataset provides AS-level edges in the Internet topology that are inferred from publicly available BGP data. Using the connectivity information from the AS relationships dataset, we build a global, undirected graph of the Internet (G). Each country's AS-level graph $C_i $ is an induced sub-graph of G. To construct each $C_i$ we find all edges in G where either the source or the target of the edge belongs to the country i.  We use these graphs as a starting point for per-country AS level topologies.
\fi

\subsection{Increasing Coverage}
\label{sec:improv_cov}

A key challenge we face in this work, is contending with known incompleteness
of existing AS-level topologies. Specifically, edges close to the network edge, 
particularly settlement-free peering edges are particularly hard to observe. Further, 
the bulk of data for Internet topology mapping is derived from BGP monitors that tend 
to be located in the Americas, Europe and Asia \vs the Middle East, Africa and other regions 
known to be implementing online information controls. We use the RIPE Atlas platform~\cite{ripeatlas} 
and perform targeted traceroutes to uncover these missing edges.

We focus on illuminating two key types of 
connectivity: (1) international connectivity of each country and (2) domestic 
connectivity of the countries. We devise two tracerouting strategies to uncover 
these two types of edges.

\myparab{Inside-out and Outside-in.} In order to find undiscovered international edges from a country, we need to traceroute from domestic sources to international destinations and vice versa (we call these traceroutes ``Inside-Out'' and ``Outside-in''). We perform traceroutes from a set of domestic RIPE Atlas probes to a randomly selected set of international probes for each country. This technique of measuring ``Inside-Out'' exposes links that are used by traffic flowing out of the country. Similarly, we perform traceroutes from the randomly selected international probes to the set of domestic probes (measuring ``Outside-In'').  The size of the subset of probes (both domestic probes and international) is progressively increased, starting from 5, in steps of 5 until no new edges are seen in 3 consecutive sets of measurements. 

\myparab{Mesh.}
  For discovering new edges local to the country (domestic edges connect two domestic ASes), we traceroute from a domestic source to a domestic destination (we call these traceroutes ``Mesh traceroutes'' since they are between all pairs of domestic traceroute vantage points, forming a mesh).

\myparab{Mapping traceroutes back to AS paths.} We take care when converting our new 
traceroute measurements to AS paths. Specifically, we use a list of prefixes belonging 
to Internet eXchange Points (IXPs)
published by PeeringDB~\cite{peeringdb} to identify and remove IP hops in our traceroutes that are located in IXPs. After removing hops in IXPs, we use CAIDA's IP prefix to ASN mapping~\cite{pfx2asn} to 
convert the traceroute to an AS-level path. 

\myparab{Inferring relationships.} Inferring business relationships between ASes is 
a challenging problem and open area of research. We take the following approach to infer 
business relationships on new edges discovered via our traceroute measurements. We begin 
by retrieving the inferred relationships for any edges that appear in our traceroute and in 
the existing dataset from CAIDA. We use these inferred relationships combined with the 
assumption that ASes will only transit traffic between two neighboring ASes if at least one of these neighbors is a customer (the ``valley free'' assumption) to constrain the business relationship for a given edge.

Table~\ref{tab:table1} summarizes the new edges found using the two methods described above.
From the table it is clear that domestic mesh traceroutes were the most beneficial, exposing a total of 5,562 new edges. Further, the benefit of adding more probes to the Inside-out traces 
begins to level off around 25 probes. Table~\ref{tab:topcountries} summarizes the countries 
that benefited most from the additional measurements and the number of new edges uncovered in each case.

\begin{table}[ht]
\centering
{\small
 
 \begin{tabular}{||c c||} 
 \hline
 Measurement Type & Total New AS edges\\
  \hline\hline
 I/O with 5 probes & 85 \\
 I/O with 10 probes & 180 \\
 I/O with 15 probes & 334 \\
 I/O with 20 probes & 509\\
 I/O with 25 probes & 647\\
 Domestic mesh & 5,562\\  %% use comma to separate thousands
 \hline
 \end{tabular}
 }
 \caption{Benefit of measurements in terms of newly discovered AS edges (I/O represents Inside-out/Outside-in traceroutes).}
 \label{tab:table1}
 \end{table}
 
 \begin{table}[ht]
\centering
{\small
 
 \begin{tabular}{||c c||} 
 \hline
 Country & Total AS edges\\
  \hline\hline
 RU & 957 \\
 US & 547 \\
 FR & 443 \\
 GB & 441 \\
 UA & 304 \\
 \hline
 \end{tabular}
 }
 \caption{Top countries in terms of new edges.}
 \label{tab:topcountries}
 \end{table}

\section{AS Topology and Internet Freedom}

We study the connection between the AS-level topology derived in the previous section and the freedom of information in a given country. The Freedom House Freedom of the Press Index (FPI), lying between 0 (high freedom) and 100 (low freedom), serves as a proxy for the freedom of information in a country. Our approach is to use features extracted from AS-level topologies as input to machine learning methods to predict $100-\textrm{FPI}$ as the target metric, which has a higher value for a higher degree of freedom.

In this section, we describe the features and machine learning methods that we use in our approach. We evaluate the accuracy of these methods in \xref{sec:evaluation}.

\subsection{Features of the AS topologies}
\label{sec:features}
For each AS-level topology, we compute a set of features and meta-information.
We break these features into four broad categories. Below, we briefly describe each class with a few 
examples; Table~\ref{tab:feature_table} enumerates the complete set of features considered.

\begin{enumerate} 
\item {\bf Structural Features.} This includes features of the 
AS-graph related to its structure such as the number of nodes or edges and connectivity features.
\item {\bf International Connectivity Features} This include features related to how a country
connects to networks in other countries. Specifically, how many countries 
and networks does it connect to and what are its path lengths to other countries.
\item {\bf IP Demographic Features.} This captures how much 
of the IP address space does the country control. It also considers the relationship between IP 
space and the population of the countries (\ie how many IPs are there per person).
\item {\bf BGP Routing Features.} This includes features related to properties of ISPs 
and networks in the country. We consider the counts of small stub networks, large providers,  
and percentiles of the customer cone size of networks in the country.
\end{enumerate}

\myparab{Preprocessing the features.} 
Since the values of different features lie in different ranges, we scale each feature using min-max scaling:
\begin{eqnarray}
\hat{X} = \frac{X - X_{min}}{X_{max} - X_{min}}
\end{eqnarray}
Where $\hat{X}$ is the scaled value for a feature with original value X. $X_{min}$ and $X_{max}$ are the minimum and maximum values of that feature across all countries.
After scaling, all features lie in the range (0,1). However we observe a number of outliers in the feature values. To mitigate their effect on our prediction we removed the countries with outlier features from the training data. As a result, we removed US, RU, SC and NL from the training data. The AS graphs of RU and US are much larger than the other countries. In case of NL, the presence of IXPs biases features relating to AS relationships (like the number of peering edges). SC has an IP density value magnitudes higher than all other countries.

\begin{table*}[ht]  %% t almost always t  no h!
 \centering
 \caption{Feature categories and description} 
 \vspace{-0.3cm} %% pg : this is a trick to suck the table up to the caption
 \label{tab:feature_table}
 {\small
%\begin{tabular}{ |l|l|l| }
\begin{tabular}{|p{3cm}|p{4cm}|p{8cm}|}
\hline
\multicolumn{3}{ |c| }{Features of Countries} \\
\hline
Category & Name & Description \\ \hline
\multirow{12}{*}{\parbox{3cm}{\centering Structural Features}} & $f_1$: num\_nodes & Number of nodes in the country's graph \\
 & $f_2$: num\_edges & Number of edges in the country's graph \\
 & $f_3$: percentile\_degree & $95^{th}$ percentile of the node degrees in the graph \\
 & $f_4$: diameter & Diameter of the country's AS graph \\
 & $f_5$:avg\_h\_im & Average horizontal imbalance\\
 & $f_{6}$: max\_load\_cen & Maximum load centrality of a node in the AS graph \\
& $f_{7}$: avg\_clustering & Average clustering coefficient of the graph \\
& $f_{8}$: graph\_clique\_number & Size of the largest clique in the graph \\
& $f_{9}$: alg\_conn& AUC of decay of algebraic connectivity as nodes are removed in order of AS rank \\
& $f_{10}$: frac\_conn& AUC of decay of fraction of largest connected component as nodes are removed in order of AS rank \\
& $f_{11}$: transitivity & the fraction of all possible triangles present in the graph\\
 & $f_{12}$: num\_large\_nodes & Number of nodes in the graph with degree \\ \hline
\multirow{3}{*}{\parbox{3cm}{\centering International Connectivity Features}} & $f_{12}$: max\_path\_len & maximum length of routed paths from a given country to all other countries \\
 & $f_{13}$: num\_intl\_countries & Number of countries, a country directly connects to \\
 & $f_{14}$: num\_intl\_nodes & Number of nodes providing international connectivity \\ \hline
\multirow{2}{*}{\parbox{3cm}{\centering IP Demographic Features}} &$f_{15}$: ip\_density & Number of IPs per person \\
 & $f_{16}$: num\_announced\_ip & Number of prefixes announced by the country \\
\hline
\multirow{4}{*}{\parbox{3cm}{\centering BGP Routing Features}} &$f_{17}$: num\_large\_providers& Number of ASes with customer cone size \textgreater 100\\
& $f_{18}$: percentile\_cust\_cone  & 95th percentile of the customer cone sizes in the country\\ 
& $f_{19}$: stub\_ases & Number of stub ASes \\
& $f_{20}$: tot\_peer\_edges & Number of AS edges in the graph that are p2p \\
\hline
\end{tabular}
}
\end{table*}

\subsection{Predicting Freedom of the Press Index (FPI)}
\label{sec:ml}
We predict the FPI for 170 countries\footnote{FPI values are available for 199 countries but we could not compute the feature values for some countries due to lack of information and hence these counld could not be a part of the study.} using these features.
We consider four different machine learning models of increasing complexity for this task. We note that our methods automatically account for the features that do not individually correlate with FPI.

\myparab{Linear Regression (LR).} We used the features to train a linear regression model. 
This method finds a linear function of the features that best 
predicts the FPI. This type of method performs well when the target variable is a globally linear function of the features.

\myparab{Regularised Linear Regression (LASSO).} Due to the small number of data points, the LR model may overfit in the face of many features. To avoid overfitting, we make use of LASSO (Least Absolute Shrinkage and Selection Operator) that shrinks dimensionality by selecting the most predictive features trying to jointly minimize prediction loss as well as L1-norm of the feature space.

In addition to the linear models, we consider decision tree-based approaches. The advantage 
of these approaches is that they are non-linear and divide the dataset into groups that are similar 
with respect to the features. The prediction function within each group is then decided as part of the model design.

\myparab{Decision Tree with simple averaging at the leaves (DTLA)} 
We train a decision tree model that groups similar countries in a group. We predict the FPI of countries in
the same bucket by taking an average of actual FPI values in the group.

\myparab{Decision Tree with Linear Regression at leaves (DTLR)} This model is similar to the last 
one. However, instead of predicting the FPI for a group based on its average, a linear 
regression is run within each group to predict the FPI values. We note that we only perform the linear 
regression in groups that have size at least 10; for the remaining groups we 
take the average (as in the previous approach). The prediction function within each leaf is then decided as part of the model design

\subsection{Evaluating Predicted FPIs}
\label{sec:evaluation}
We evaluate the four models described above using the 170 countries present 
in the 2012 FPI values for which it was possible to compute the features described in Table \ref{tab:feature_table}. 
We use leave-one-out cross validation (LOOCV) to evaluate the ML methods. We train each of the 4 models (LR, LASSO, DTLA, DTLR) with all but one country and then evaluate how well we predict the left\-out country. This process is repeated in turn for all 170 countries. 

\begin{figure}[t]
    \includegraphics[width=0.49\textwidth, trim={0cm 0cm 0cm 0cm},clip]{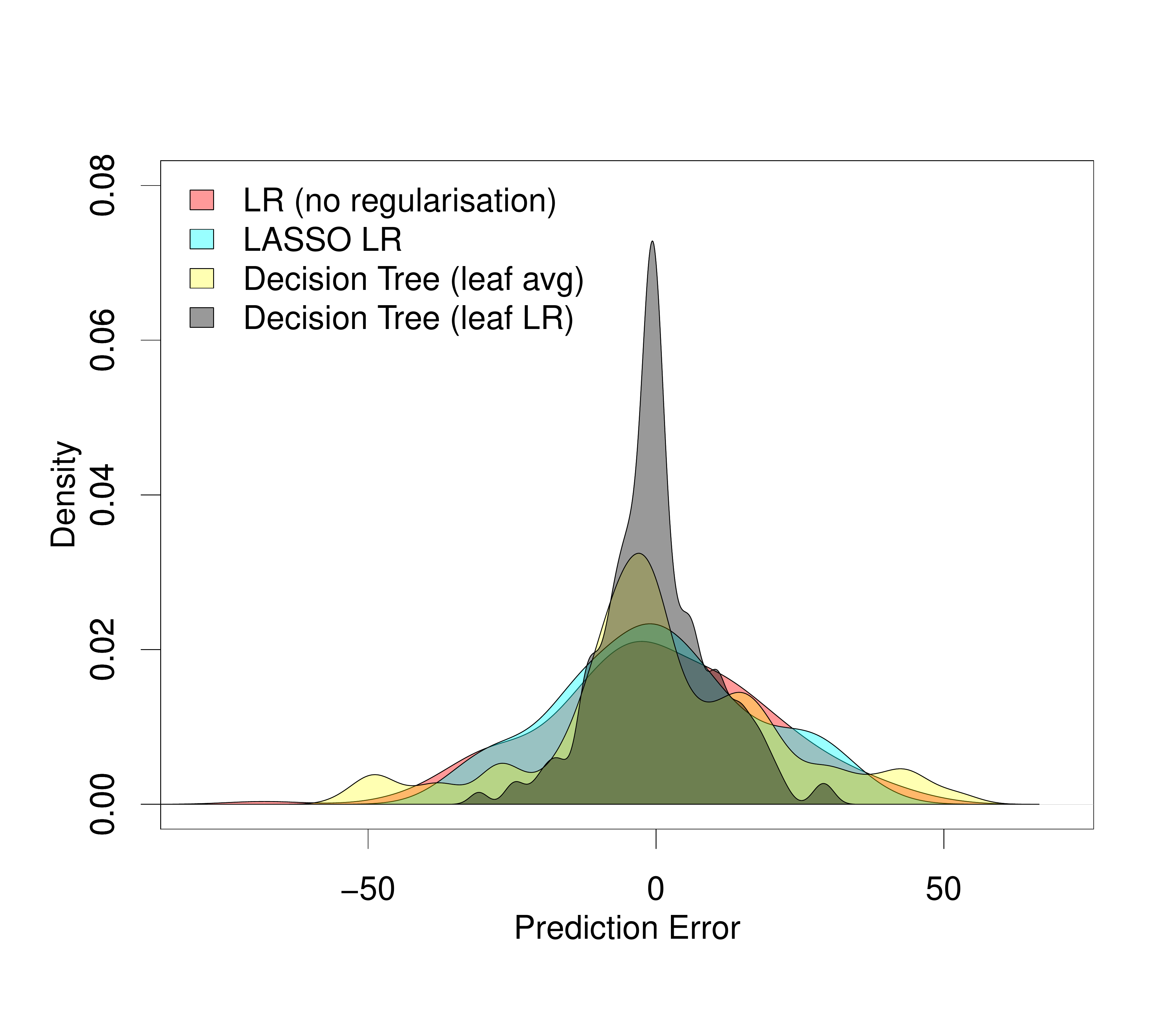}
    \caption{Comparison of the prediction error of our learning approaches.}
    \label{fig:kde_comp}
\end{figure}

\myparab{Predicting FPI.} Figure~\ref{fig:kde_comp} shows the estimated kernel density function for the distribution of prediction errors 
for each of the four aforementioned models. We find LR and LASSO that perform 
linear regression across the entire dataset, provide poor accuracy with an average prediction 
error of 15.05\%. The decision tree model (DTLA) has modestly increased 
accuracy but the average prediction error reamined close to 15.04\%. Performing linear regression within the groups produced 
by the decision tree (DTLR) has the highest accuracy with an average prediction error of 7.06\%.

\begin{figure}[t]
    \includegraphics[width=0.49\textwidth, trim={0cm 0cm 0cm 0cm},clip]{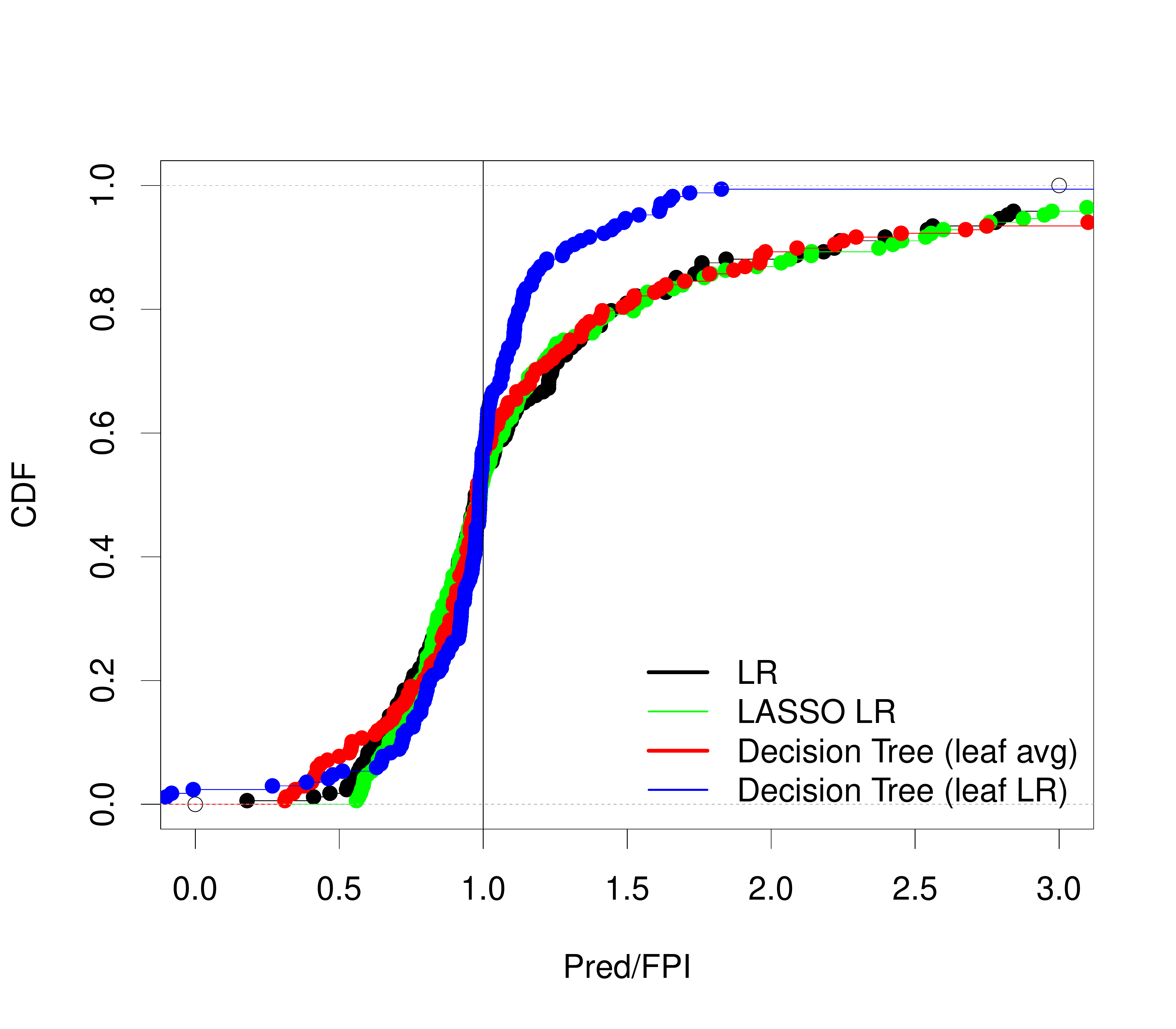}
    \caption{CDF of the prediction error of our learning approaches.}
    \label{fig:error_comp}
\end{figure}

Figure \ref{fig:error_comp} shows the cumulative distributon of normalized error (\ie $\frac{predicted FPI}{actual FPI}$) for the four models. We note that the best model--decision tree with linear regression--has an error of atmost 10\%, 71\% of the time. 

\myparab{Predicting freedom category.} Freedom House groups countries based on their FPI value: ($61-100$) Free, ($31-60$) Partly-Free, and ($0-30$) Not Free. We also consider how well our model predicts the category a country will fall under. By discretizing the predictions of DTLR, we were able to predict the freedom categories with 81\% accuracy. The countries for which DTLR predicted freedom category wrongly, often, our prediction and the actual freedom category were partly-free and free (and vice versa). If we consider free and partly-free as the same label, our accuracy of prediction improves to 91\%.
\begin{figure*}[ht]
     \includegraphics[width=\textwidth, trim={0cm 0cm 0cm 0cm},clip]{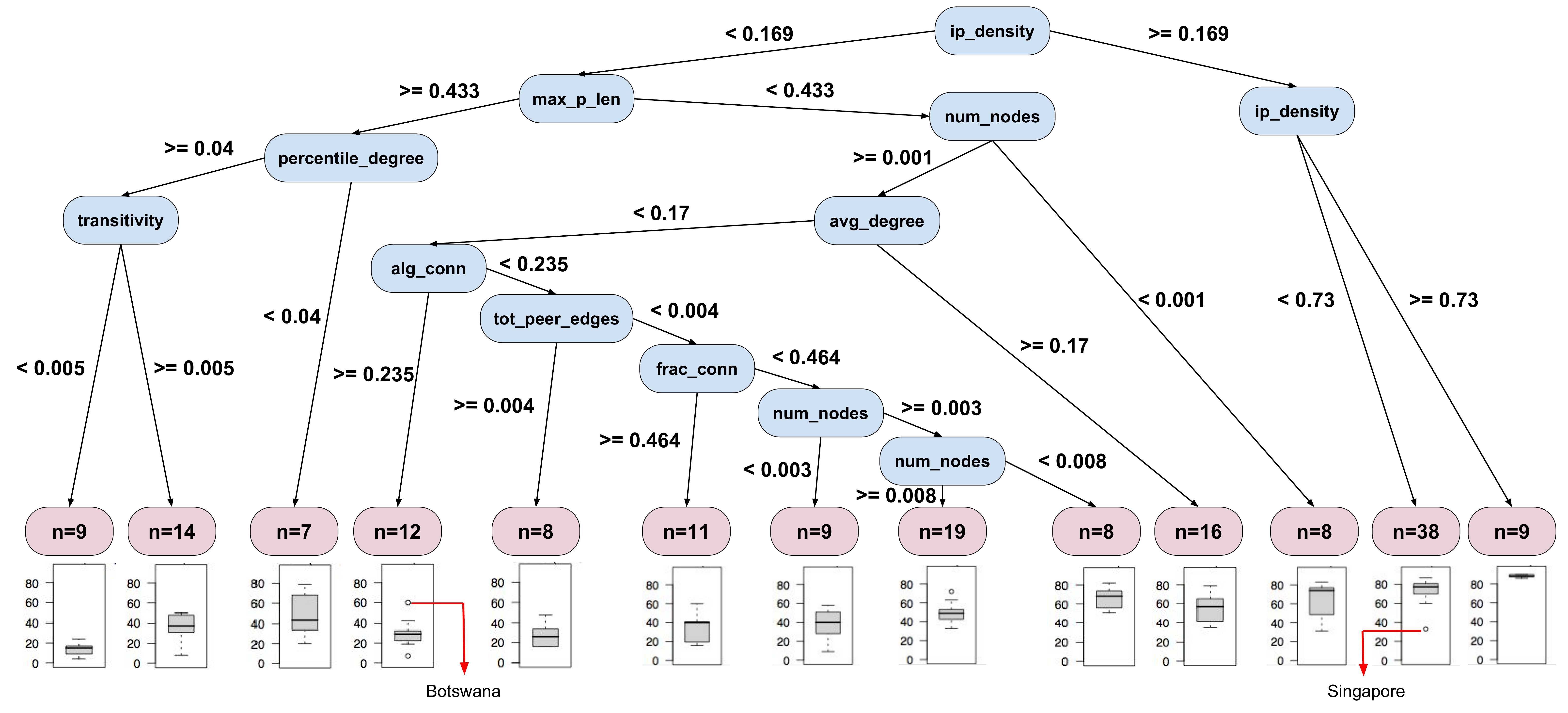}
    \caption{Comparison of the prediction error of our learning approaches. Arrow point to local outliers, see Section \ref{sec:outliers} for details}
    \label{fig:dtree}
\end{figure*} 

\section{Features that Predict Freedom}
\label{sec:feature_influence}

We now discuss which features are the most relevant 
for predicting FPI using our most accurate classification model 
(decision tree with linear regression).

IP density has the highest influence on the freedom index. As Figure \ref{fig:dtree} shows, a normalised IP density value of 0.169 or higher implies high freedom of expression in a country.
This metric captures the ratio of IP addresses to users 
within a country and can be seen as approximating 
the level of connectivity per capita in the country.

We also observe a negative correlation between the maximum length of BGP policy compliant paths from a country to all other countries (max\_p\_len). Normalised max\_p\_len value of 0.433 or lower ensures high freedom in a country. This makes intuitive sense since longer paths imply poor connectivity. 

We find that poor connectivity properties tend to correspond 
to countries with low FPI values. Countries with high path length,  low degree values and low transivity (\ie number of ``triangles'' in the graph) are among the lowest in terms of FPI scores. This first group includes countries that are known 
to implement strong information controls (\eg Ethiopia, China, Cuba).

%is indicative of poor freedom in a country. In a broader sense, more developed Internet infrastructure implies higher freedom. While this seems to be the general case, we see a violation in the case of Singapore where the Internet infrastructure is sophisticated but there the freedom index is low. We discuss such cases in the next section (\xref{sec:outliers}).

\begin{figure}[t]
\centering
\includegraphics[width=0.4\textwidth]{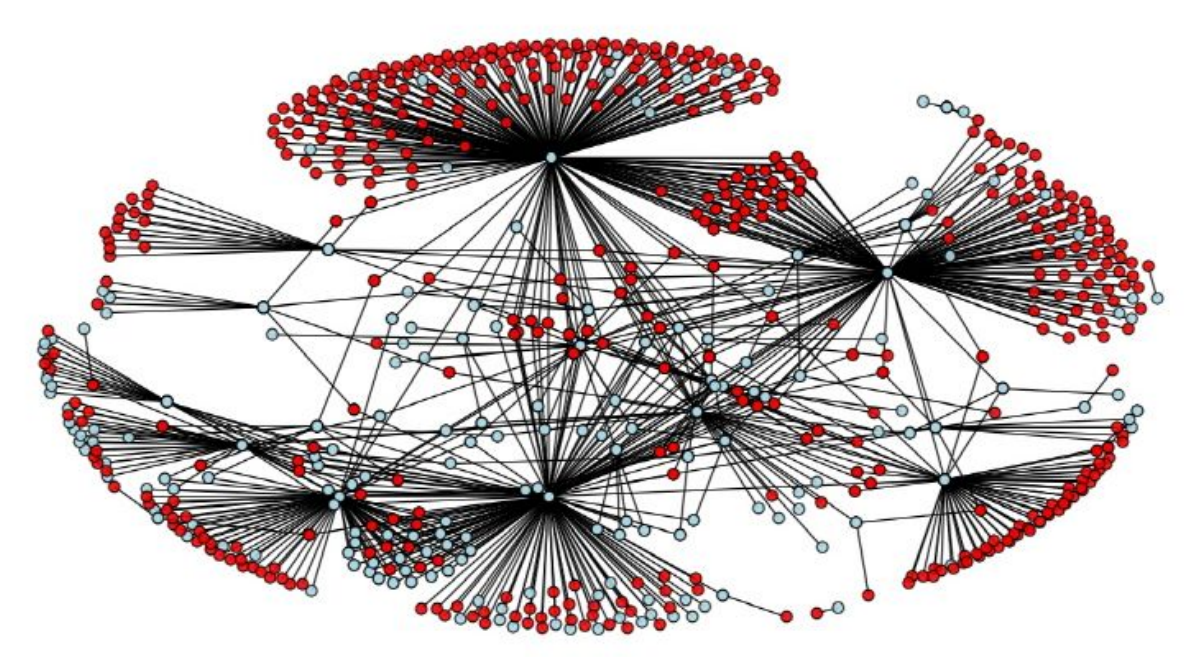}\\
{\small (a) Singapore } \\
\includegraphics[width=0.4\textwidth]{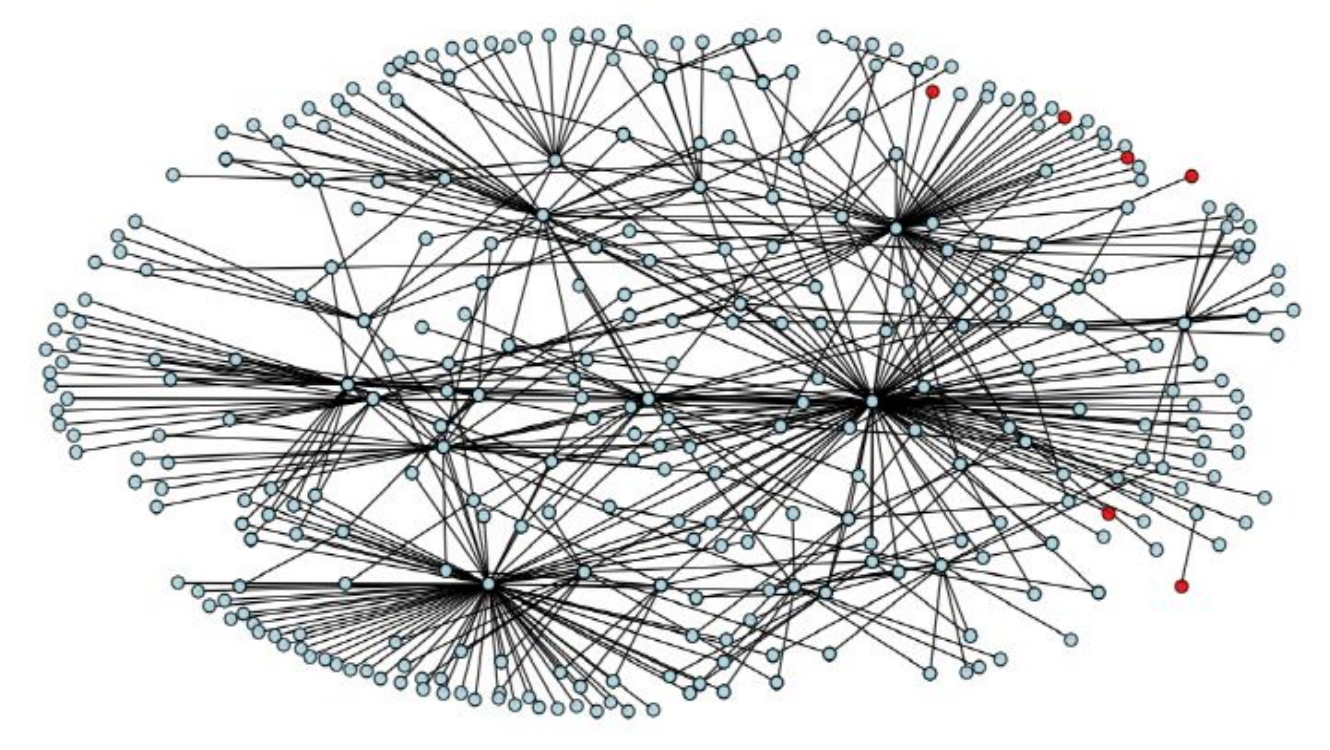}\\
{\small (b) Iran} \\
\caption{AS graphs for Singapore and Iran (red/dark 
nodes are international ASes and light nodes are domestic.}
\label{fig:irsg}
\end{figure}

\section{Identifying Unusual Countries}
\label{sec:outliers}

Our decision tree can also highlight countries with 
connectivity profiles that are not consistent 
with their information control policies. 
There two such instances that stand out in Figure~\ref{fig:dtree}: Botswana and Singapore. In case of Singapore, according to our prediction the FPI should be very high ($\>80$). But in reality the FPI index of SG is 31.
On the other hand, our FPI prediction for Botswana is very low (29.25) but in reality Botswana has high freedom of expression.

We dig deeper into the case of Singapore. Singapore 
respresents a country with a well established IT infrastructure that also implements online information controls \cite{onisg}. We can see 
this difference qualitatively in Figure~\ref{fig:irsg} 
which compares the connectivity graphs of Singapore and one 
of the least free countries, Iran. Iran shows strong limits 
in terms of international connectivity, connecting to only three international networks. Singapore, in contrast, has a rich international connectivity with 257 domestic ASes connecting to a total of 3022 international ASes. 

\section{Conclusions}
Freedom House FPI assesses the degree of freedom in digital and print media for countries across the globe. Using FPI as a measure of freedom of expression, we investigate the relationship 
between Internet infrastructure and information freedom 
around the globe. Our techniques can help bootstrap 
understandings of information freedom when empirical data 
may not be readily available. We are also able to identify 
features of AS topologies that are more representative of countries that implement online information controls. 

\myparab{Future work.} While this work presents a first exploration of the relationship 
between Internet infrasturcture and information freedom, 
there is still much ground to be covered in this space. 
In future work, we plan to take a two pronged approach 
to extend this study. Specifically, we hope to leverage 
social science expertise 
to better reason about the social and political factors 
that impact information policy and discuss our findings with operators of existing large networks to see how policy 
shapes their day-to-day network management. 

%developed a method to reduce the manual effort in computing this measure of freedom. We developed a method that uses publicly available demographic information (population, IPs allocated) for countries and empirically inferred structure of the country's AS network. Our per country FPI predictions differ from the Freedom House values by only 7.5\% on an average. We also predict the freedom labels (free, not free) labels for a country with 91\% accuracy. The decision tree for FPI prediction can be read and its decisions make intuitive sense. Our study highlights characteristics of countries that correlate with freedom.
%\ifnum\anon=0
%	\section*{Acknowledgments}
%\fi

%% not losing space to refs!
{\footnotesize
\bibliographystyle{abbrv}
\bibliography{bibliography}
}
\end{document}